%% file: Header.tex
\documentclass{sigchi}

\usepackage{booktabs} 
\usepackage{balance}       
\usepackage{graphics}      
\usepackage[T1]{fontenc}   
\usepackage{txfonts}
\usepackage{mathptmx}
\usepackage[pdflang={en-US},pdftex]{hyperref}
\usepackage{color}
\usepackage{booktabs}
\usepackage{textcomp}
\usepackage{multirow}
\usepackage{enumitem}
\usepackage{cite}

\usepackage{microtype}        
\usepackage{ccicons}          

\def\plaintitle{Chasing Lions: Co-Designing \\Human-Drone Interaction in Sub-Saharan Africa}
\def\plainauthor{Anna Wojciechowska, Foad Hamidi, Andres Lucero, Jessica R. Cauchard}
\def\plainkeywords{Human-Drone Interaction, Co-Design, HCI4D, Social Robotics, Drone, UAV, Design.}

\makeatletter
\def\url@leostyle{%
  \@ifundefined{selectfont}{
    \def\UrlFont{\sf}
  }{
    \def\UrlFont{\small\bf\ttfamily}
  }}
\makeatother
\def\pprw{8.5in}
\def\pprh{11in}

\definecolor{linkColor}{RGB}{6,125,233}
\hypersetup{%
  pdftitle={\plaintitle},
  pdfauthor={\plainauthor},
  pdfkeywords={\plainkeywords},
  pdfdisplaydoctitle=true, 
  bookmarksnumbered,
  pdfstartview={FitH},
  colorlinks,
  citecolor=black,
  filecolor=black,
  linkcolor=black,
  urlcolor=linkColor,
  breaklinks=true,
  hypertexnames=false
}

\teaser{ 
\centering
    \includegraphics[height=5.65cm,keepaspectratio]{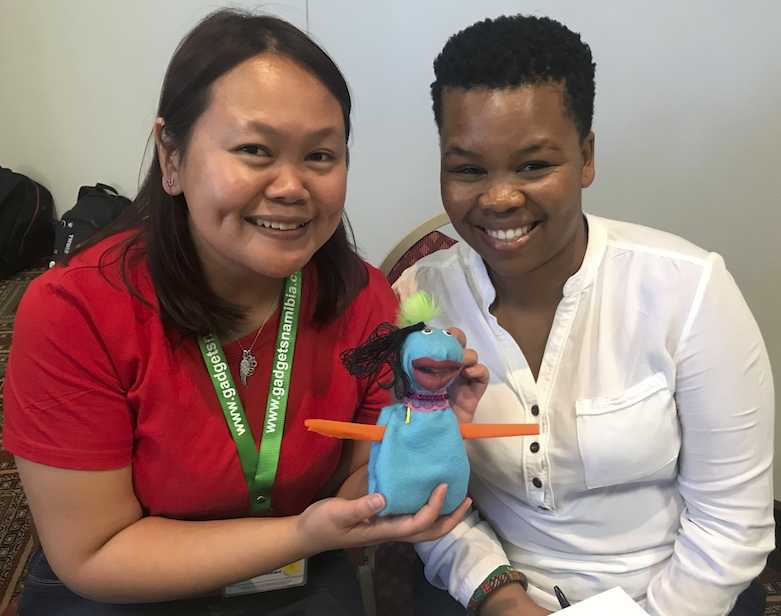}
    \includegraphics[height=5.65cm,keepaspectratio]{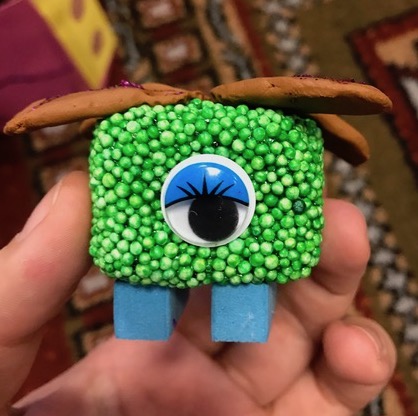}
    \includegraphics[height=5.65cm,keepaspectratio]{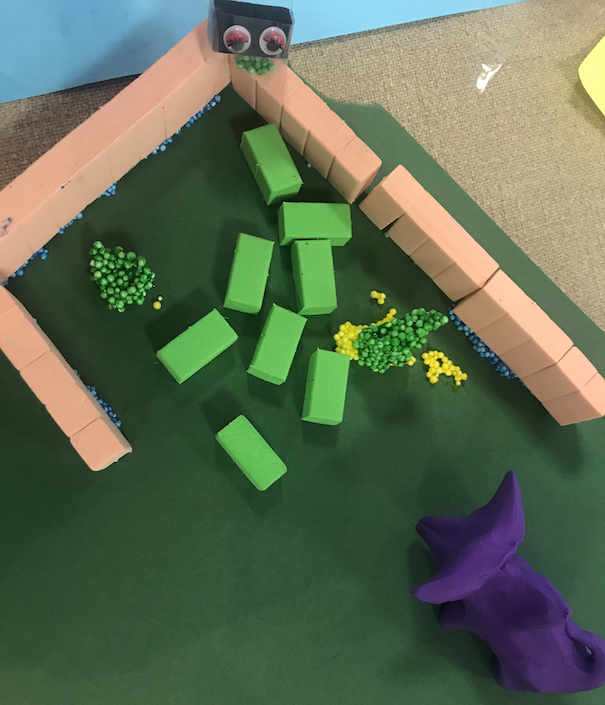}
\caption{Example of drones resulting from the co-design activity. (Left) Participants posing with their \textit{Learning} drone ``Teacha Lola''. (Center) \textit{Agriculture} drone ``Jo-Ja'', a bug that detects other bugs in fields. (Right) Scene of a drone monitoring cattle against predator attacks.} 
\label{fig-teaser} 
}

\toappear{\scriptsize Permission to make digital or hard copies of all or part of this work for personal or classroom use is granted without fee provided that copies are not made or distributed for profit or commercial advantage and that copies bear this notice and the full citation on the first page. Copyrights for components of this work owned by others than the author(s) must be honored. Abstracting with credit is permitted. To copy otherwise, or republish, to post on servers or to redistribute to lists, requires prior specific permission and/or a fee. Request permissions from permissions@acm.org. \\
{\emph{DIS '20, July 6--10, 2020, Eindhoven, Netherlands.} } \\
\copyright~2020 Copyright is held by the owner/author(s). Publication rights licensed to ACM. \\
ACM ISBN 978-1-4503-6974-9/20/07\ ...\$15.00.\\
http://dx.doi.org/10.1145/3357236.3395481}

\begin{document}

\title{\plaintitle}

\numberofauthors{1}
\author{%
  \alignauthor{Anna Wojciechowska$^{1,2}$, Foad Hamidi$^{3}$, Andr\'es Lucero$^{4}$, Jessica R. Cauchard$^{1,2}$\\
    \affaddr{$^{1}$Magic Lab, Ben Gurion University of the Negev, Be'er Sheva, Israel}\\
    \affaddr{$^{2}$Interdisciplinary Center (IDC) Herzliya, Herzliya, Israel}\\
    \affaddr{$^{3}$University of Maryland, Baltimore County, Baltimore, Maryland, United States}\\
    \affaddr{$^{4}$Aalto University, Espoo, Finland}\\
    \email {wojciechowska.anna2511@gmail.com, foadhamidi@umbc.edu, \{lucero, jcauchard\}@acm.org}\\}
    }
    
\maketitle

\begin{abstract}
Drones are an exciting technology that is quickly being adopted in the global consumer market. Africa has become a center of deployment with the first drone airport established in Rwanda and drones currently being used for applications such as medical deliveries, agriculture, and wildlife monitoring. Despite this increasing presence of drones, there is a lack of research on stakeholders' perspectives from this region. We ran a human-drone interaction user study (N=15) with experts from several sub-Saharan countries using a co-design methodology. Participants described novel applications and identified important design aspects for the integration of drones in this context. Our results highlight the potential of drones to address real world problems, the need for them to be culturally situated, and the importance of considering the social aspects of their interaction with humans. This research highlights the need for diverse perspectives in the human-drone interaction design process.
\end{abstract}


%
%

\ccsdesc[500]{Human-centered computing~HCI design and evaluation methods}
\ccsdesc[500]{Human-centered computing~Interaction design}
\ccsdesc[300]{Human-centered computing~Collaborative and social computing design and evaluation methods}
\ccsdesc[300]{Social and professional topics~Cultural characteristics}

\keywords{\plainkeywords}

\printccsdesc

\input{body.tex}

\end{document}

%% file: body.tex

\section{Introduction}
Drones are an exciting emerging technology with a fast growing adoption in the global consumer market, where small-sized drones are being sold worldwide from airport duty free shops to supermarket aisles and toy stores. Their rapid expansion is predicted to reach 29 million shipments by 2021 \cite{Joshi2017CommercialForecast}. These flying robots present new applications to support users in a mobile context. They also present unique opportunities from the Human-Computer Interaction perspective and have now their own sub-field of Human-Drone Interaction research \cite{Cauchard2015DroneInteraction}.

Drones present many usages and applications from photography to journalism, agriculture, monitoring, and scientific work. They are particularly useful when deployed in locations where infrastructures are not reliable, such as over large land and sea areas, after disasters, and in applications in low and middle income countries. In such environments, these flying machines can help gather data and support people on the ground without the need for complex and expensive infrastructure. Rwanda appears to be the first country in the world to have built a ``drone port'' for civilian and commercial usage \cite{Kuo2015TheAfrica}, bringing the Global South to the center of this technological development. On the African continent, drones are being used for a variety of applications, such as delivering medical supplies to rural areas \cite{Mendelow2002DevelopmentLaboratories}, mitigating conflicts between humans and wildlife \cite{Hahn2017UnmannedStudy}, and as an anti-poaching measure \cite{Penny2019UsingTactic}. Sandvik argues that the \textit{`African drone' has
become a vehicle for the production and distribution of norms, resources,
and forms of legitimacy that have implications for drone proliferation, both
within and outside Africa''}~\cite{Sandvik2017}.

Despite their fast adoption on the African continent, drones are primarily designed and manufactured elsewhere (e.g., China, USA). We argue that the long lasting adoption and positive impact of such technology, especially as it gains increasing autonomy, will depend on its acceptability by the population confronted with it and the relevance to its context of deployment. It is therefore crucial to use participatory approaches to understand and incorporate African perspectives in drones' design and development. This research focuses on the timely need to include future users in the drone design process. 
This paper presents results from a co-design workshop conducted with 15 sub-Saharan experts exploring future applications and feature design for drones in the region. Our results highlight several important topics around drone usages, designs, and expectations; including several surprising findings, like the desire to use drones to protect people against animal attacks. 

This work's contributions are as follows:

\begin{itemize} [noitemsep]
    \item Research methodology for co-designing drones
    \item Findings from a drone study with sub-Saharan experts
    \item Novel applications for human-drone interaction
    \item Factors for introducing drones in sub-Saharan Africa
\end{itemize}
    
To the best of our knowledge, this work is the first drone co-design study; and the first of its kind in sub-Saharan Africa. Our overarching goal is to diversify the research field and design space, ensuring that people who will be directly impacted by this emerging technology have a voice in this space. Our results highlight large cultural diversity and creativity that would not have been exposed without involving this population. 

\section{Related Work}
This section starts with prior work in human-drone interaction, then discusses co-design as a methodology and how co-design has been previously applied in the sub-Saharan context.

\subsection{Human-Drone Interaction}
Human-Drone Interaction (HDI) is a growing research area within the Human-Computer and Human-Robot Interaction (HCI/HRI) communities. In a recent survey paper \cite{Mirri2019Human-DroneChallenges}, specific research challenges were identified, including how drones can be integrated in people's daily life and how to understand one's perception of drones. Interestingly, these are some of the main issues tackled in this work. Prior works have investigated drones used in daily activities, e.g., jogging \cite{Mueller2015JoggingQuadcopter}, navigational aid \cite{AvilaSoto2017,Knierim2018Quadcopter-ProjectedAwareness}, tour guide \cite{BrockAnkeM.2018,Cauchard2019Drone.io:Interaction}, for dance performance \cite{Kim2018Aeroquake:Dance,Eriksson2019DancingIntercorporeality}, and even games \cite{Kljun2015StreetGamez}. Wojciechowska et al. \cite{Wojciechowska2019} recently presented a model of perception based on drones' physical properties. Prior works had investigated characteristics to make a \textit{social drone} suitable for interaction and companionship. Kim et al. \cite{Kim2016TheDevice} suggested the use of ``adorability'' features, while other works proposed a blue oval shaped drone fitted with a tablet to display a ``friendly face'' \cite{Yeh2017} or a round shape drone with a face \cite{Karjalainen2017SocialEnvironment}. In addition to physical features, prior works explored affect and emotion as a step towards integrating drones into human environments. Preliminary work suggested that a drone could convey different emotional states \cite{Arroyo2014a}, which was then validated in our prior work using the drone's movements and behaviors \cite{Cauchard2016b}. These prior works show that appropriate design is intrinsic to integrating drones into human spaces. 

In terms of methodology, in a 2019 survey \cite{Baytas2019TheEnvironments}, Bayta\c{s} et al showed that only four prior works had reported using a design workshop methodology in HDI. Out of these works, two were used to design social drones \cite{Karjalainen2017SocialEnvironment,Yeh2017} which resulted in an understanding of social drone physical features and potential tasks. While these works improved our understanding of the look and feel of a social drone, they offer limited findings into the actual integration of such technology in human spaces. We here opted for a co-design methodology to integrate diverse perspectives in the design process, as described below.

\subsection{Co-Design Methodology}
Practitioners from different research and design fields have understood the importance of involving diverse groups of users in the generation phase of novel artefacts, products, and services; thus facilitating participation has become one of the cornerstones of co-design \cite{brandt2007tangible}. Underpinning this approach is the supposition that stakeholders, including users, can contribute productively through involvement in the design process as they bring privileged insights into the domain that designers are trying to address; and in the ways in which future products and services may fit into and affect that domain. Within this field, there exists a variety of methods, techniques, and events intended to inspire design participants and scaffold collaborative ideation and concept development, such as dialogue-labs.

Dialogue-labs \cite{Lucero2012} are primarily used in the middle stages of the design process to support researchers and designers in creating ideas and concepts for future designs, together with relevant stakeholders and end users. It combines different methods and techniques, providing a structured way of generating ideas through a sequence of co-design activities. Its main focus is on three key structuring aspects: the process of how dialogue-labs sessions are orchestrated, the space in which the sessions unfold, and the materials that are employed. Dialogue-labs have been successfully applied in a wide range of projects to involve different participants including HCI researchers \cite{Goguey2019}, children with ADHD \cite{fekete2019p}, and animals \cite{Hirskyj-Douglas2019}.

Co-design approaches have successfully brought under-represented users perspectives to the design table. The section below describes prior work in the African context. 

\subsection{Co-Design in Sub-Saharan Africa}
A growing body of work is recognizing the importance of using participatory and community-based methodologies for conducting HCI research in the African context \cite{Hamidi2018, Winschiers-Theophilus2010BeingApproach,Molapo:2016,Dell2016TheDevelopment}. Part of it is rooted in earlier efforts of the Information and Communication Technologies for Development community and more recently, HCI4D, defined \cite{Ho2009Human-ComputerFuture} as: the study of \textit{``how interactive products, applications, and systems can be appropriately designed to both address the distinctive needs of users in developing regions, and to cope with the difficult infrastructural contexts where these technologies must work.''} In a survey of HCI4D research, Dell and Kumar outlined the importance of recognizing the equal contribution that work in this space can offer to the broader HCI community \cite{Dell2016TheDevelopment}. They observed that ``\textit{there is a lesser focus now on designing new hardware for users in developing world contexts with the growing convergence of technologies and mobile computing becoming increasingly accessible and affordable for the North and South.}'' Yet, with production concentrated in specific parts of the world (e.g., China), we argue that it is crucial to include diverse perspectives on the design of emerging technologies, including specialized hardware. These efforts will not only result in designs that are more sensitive to the needs and desires of people in these contexts, but will also, in line with Dell and Kumar's claim, contribute to the broader HCI design space. 

Previous research in sub-Saharan Africa has underlined the importance of considering the sociocultural context of design activities \cite{Winschiers-Theophilus2010BeingApproach, Hamidi2018}. Winschiers-Theophilus et al. \cite{Winschiers-Theophilus2010BeingApproach}  previously discussed how ``\textit{storytelling, inclusive decision making and participatory community meetings are key features in traditional rural African communities,}'' and how these participatory practices need to be considered when conducting co-design in this region to ``\textit{guide a closed group towards a design output.}'' A recent survey showed that prior research in the region often focused on specific user groups, such as under-served populations, or specific use cases \cite{VanBiljon2019}. Other research has recommended the inclusion of early and frequent feedback from multiple diverse stakeholder as important strategies to keep projects relevant to their end-user \cite{Hamidi2018, Molapo:2016}. 

In this project, we employed a participatory approach inspired by previous efforts, and yet different in important ways. Specifically, our participants consisted of a diverse group of stakeholders from different regions of sub-Saharan Africa. Given the participant group's diversity, we adopted our methodology to focus on broader questions about drone design and used hands-on co-design activities to ground the discussions.

\section{CO-DESIGN WORKSHOP}
\label{sec:sim}
This sections outlines our methodology that is tailored for the exploration of drones, a novel and unfamiliar technology, not yet well-established in people's minds. A member of the research team moderated and two took part in the activities.

\subsection{Context and Location}
The study was run as a formal workshop at AfriCHI 2018 \cite{CauchardAfriCHI}, the 2nd African HCI conference, organized by the Namibian University of Science and Technology (NUST) in Windhoek. The workshop was advertised on the conference website and social media, described as a hands-on workshop that brings together researchers and practitioners to discuss the cultural aspects of Human-Robot Interaction design. Participants were recruited through the conference, via the website and social media, and with the help of the conference organizers.

\subsection{Participants}
We recruited 15 participants (5f,10m) living in sub-Saharan Africa, age 18-60 y.o. ($\mu$=35 y.o.) and a minor who came with his mother who filled in a parental consent form. Seven (7) participants were Namibian, 3 Zimbabwean, 2 Nigerian, 1 Mesotho, 1 Indonesian and 1 Italian who had both been living in Namibia for many years. All participants had full proficiency in English, which is one of the official languages in many sub-Saharan African countries, and spoke other languages such as: Afrikaan, Oshiwambo, Shona, Yoruba, French, Hausa, Sesotho, Damara Nama dialect (Khoekhoe language), Portuguese, Italian, Spanish, and Swedish. Eleven participants lived in urban areas, 2 in suburban, 1 in rural, and 1 in both urban and rural areas\footnote{Many participants currently living in urban areas later described growing up in rural areas and regularly visiting family in the bush.}. Most (13) had seen a drone before, including 2 who had piloted one before and 1 drone owner.

\subsection{METHOD}
The co-design workshop was organized in seven stages.

\subsubsection{Stage 1: Welcome}
Participants were welcomed individually. An introduction was then given to the group where participants were informed that the workshop data would be collected for research purposes and that the results would be made available to them. They were then presented with the consent form.

\subsubsection{Stage 2: Pre-study Surveys}
After signing the consent form, participants were asked to fill in the following three surveys on their own.

\textbf{Values Survey Module (VSM):} 30-item questionnaire for comparing culturally influenced values and sentiments \cite{Hofstede2015}.

\textbf{Negative Attitude Towards Robots Scale (NARS):} 14-item self-report inventory measuring attitude towards robots \cite{Nomura2006MeasurementRobots}.

\textbf{Demographics:} Survey about languages spoken, countries of residence, and prior experience with drones.

\subsubsection{Stage 3: Introduction} To inspire participants about drones and how co-design would be used, two expert presentations were given. The first one focused on human-drone interaction research and the second one introduced co-design methodology and emphasized its importance with regards to the idea creation process.

\subsubsection{Stage 4: Focus Group} After a short break, participants were invited to sit in a circle for the focus group which lasted approximately 1.5 hours. It was facilitated by the moderator and was audio and videotaped. The session started with a short introduction of the participants as an ice-breaker. Participants were then asked a series of questions regarding social drone characteristics, current as well as future applications, acceptable control mechanisms, and interaction modalities. These questions were used to stimulate discussions. At the end, all discussed applications were written down on a flip chart. Each participant had three stickers to vote for their favorite application, which would then be used for the design activity. During the focus group, photos of drones were projected on the wall to inspire participants.

\subsubsection{Stage 5: Design Activity}
The top five applications were associated with a station, each arranged with a different set of craft material and some inspiration props to allow participants to build a common design language and provide them with different entry points to the design problem \cite{Lucero2012}. The stations were arranged as follows:

\textbf{Agriculture - Modelling:} Clay, foam blocks, and foam clay. The props used for the station were Virtual Reality cardboard glasses showing a 360 video drone footage.

\textbf{Healthcare - Collage:} Magazine clippings, specialty papers, and stickers. 
The props used for the station were drone concept videos showed on a tablet. 

\textbf{Security - Lego:} Lego blocks with the exclusion of human and animal figures. The props used for the station where two Pocket Robots that could dance and record audio messages.

\textbf{Learning - Fabric:} Felt, velvet, crepe paper, fringe trims, sequins, feathers, strings, needles, staples, glue, and scissors. The prop used for the station was a small-size robot.

\textbf{Entertainment - Sketching:} Various types of paper, pens, crayons, markers, pastels, highlighters, and glitter.
The prop used was a small-size drone (Parrot Mambo) with safety guards that participants could control through a companion phone and fly around the room as they wished.

\begin{figure}[t]
\includegraphics[width=1\columnwidth]{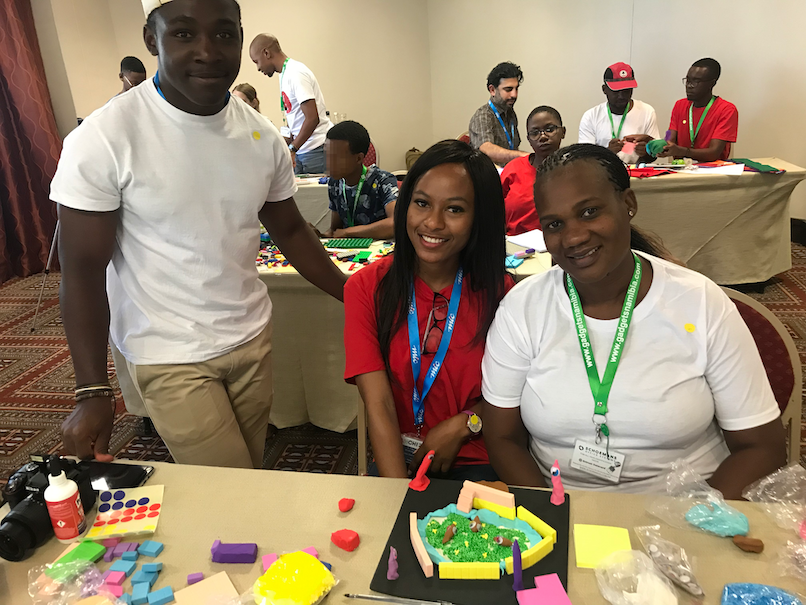}
\caption {Participants posing with their \textit{Agriculture} drone}
\label{fig:activity}
\end{figure}

After lunch, participants were split into small groups of two to three participants based on the dynamics of the focus group. They were asked to portray a drone for each chosen application (Figure \ref{fig:activity}), including but not restricted to: form -- look \& feel --, interaction modality, sociability, and capabilities. The groups were given 15 minutes per station to create their design, with a total of four stations per group.

\subsubsection{Stage 6: Idea Presentation}
Each group presented their visual representations and detailed the concept and motivation for their designs, one application at a time. Participants voted for their favorite drone design in each application and filled in a questionnaire to assess the quality of the voted design by rating them on a seven-point Likert scale (where -3 is `very bad' and 3 is `very good'). 

\subsubsection{Stage 7: Wrap Up}
The research team thanked the participants for their time and participation. Participants were then offered the opportunity to stay in touch for further collaboration in the project.

\begin{table*}[t]
\centering
\begin{tabular}{p{2cm}|p{4cm}|p{10.5cm}}
Type & Name & Description \\
\hline
\hline
    \multirow{7}{*}{Look} & Anthropo/Zoomorphism & Design inspired by a person, or an animal or an insect. \\
    \cline{2-3}
     & Eyes & $\geqslant$ 1 feature can be understood as one or multiple eyes.\\
    \cline{2-3}
     & Mouth & $\geqslant$ 1 feature can be understood as the shape of a mouth, snout, or beak.\\
    \cline{2-3}
     & Facial Features & $\geqslant$ 1 feature can be understood as a facial feature.\\
     \cline{2-3}
     & Limbs & $\geqslant$ 1 feature can be understood as a limb (e.g., hands, legs, wings).\\
     \cline{2-3}
     & Shape & The drone's body is designed primarily in a \textit{Round} or in a \textit{Cubic} shape.\\ 
\hline
    \multirow{4}{*}{Capabilities} & Photo/Video & The drone is equipped with a camera. \\
    \cline{2-3}
     & Aerial Features & $\geqslant$ 1 feature is for flying purposes, such as wings or propellers.\\
    \cline{2-3}
    & Aquatic Features & $\geqslant$ 1 feature is for swimming purposes, such as fins or tentacles.\\
    \cline{2-3}
    & Ground Features & $\geqslant$ 1 feature is for ground movement, such as wheels or legs.\\
\hline
    Interaction & Feedback & The feedback is designed for \textit{Audio}, \textit{Visual}, or \textit{No feedback}. \\
\hline
\end{tabular}
\caption{List of features used to code the drones that were created during the Design Activity, inspired by \cite{Wojciechowska2019}.} 
\label{tab:droneCoding}
\end{table*}

\section{Data Analysis}
We analyzed the focus group, drone designs, and idea presentations. The NARS and VSM questionnaires are not analyzed at this point. This section describes how the data was analyzed.

\subsection{Focus Group Analysis}
The audio recordings were transcribed by a third-party company. A member of the research team then listened to the audio files while reading the transcripts to improve their overall quality and ensure that technical terms were properly identified. The low quality of the audio recording did not allow the identification of specific speakers. Two members of the research team independently conducted thematic analysis of the data using inductive analysis and identified six different themes. The themes were then emailed to some of the study participants interested in giving further feedback for validation. 

\subsection{Design Activity Analysis}
The designs were analysed using features based upon prior work \cite{Wojciechowska2019}, the data itself, and discussions among the research team. Six features are related to the drone's look, four define capabilities, and one interaction (Table \ref{tab:droneCoding}). The recordings from the Idea Presentation stage were transcribed by one of the researchers. The information gathered in these audio and video files is used conjointly with the researchers' notes to understand how the drones were designed. The Idea Presentation concluded with 13 participants voting on the ``best design'' for which we computed the average and standard deviation score. 

\section{Results}
This section presents the results of the workshop divided in the findings from the focus group and the design activity, as well as the final ratings of the ``best drone designs''.

\subsection{Focus Group Findings}
Our analysis led to six themes that we present below. 
\subsubsection{Drones Embodying Values as Extensions of Communities}

A major theme that was present throughout the discussion involved how the form and function of a drone would and should reflect the values practiced in a community. In this sense, the drone would be viewed as an extension of the community.

Several participants described how safety was of utmost importance to their communities. For these participants drones could help protect different areas for the community:

``\textit{It could be just for the certain area and maybe connected to the residents of, uh, of that area ... And also to the control centre of the law enforcement.}''

Other participants differentiated between the safety needs of urban and rural communities: ``\textit{if you look at urban areas, people are looking at the robberies, muggings, and things like that. And in rural areas, it's more to do with women who are raped, or, people coming to steal animals.}'' Other participants described hospitality and friendliness as contrasting values that could be reflected by a drone: ``\textit{drone .... is a representation of you as a community. ... If you are a friendly community and there's a stranger [you might send them a drone] with a water gun or whatever [laughter].}''

Reflecting on these contrasts, another participant mentioned that drones could reflect values and moral principles represented by a specific community: 

``\textit{It could be an extension of the community.... So some of the values of the different communities maybe could be ... implemented into the drone. For some communities, the value of privacy might be very important. So we want to make sure coming to that small community, you can encounter that. Whereas in another community maybe the value of education or art expression might be [important].}''

As they represent specific morals and beliefs, drones could assist with inter-communities communication: 

``\textit{We were talking earlier on having different communities and each community could have a drone, and they could network.}''

\subsubsection{Regulation, Culture and Innovation}

Another highly discussed theme described a tension between a desire for innovation, the need for regulation and the importance of acknowledging cultural norms and traditions, especially in rural communities. While most of the participants were excited about the possibilities of novel drone applications, the majority of participants also described the importance of some form of introductory process where technology is mindfully brought into communities, warning against making assumptions about the similarity of different contexts. 

For example, one of the participants who described his community as \textit{``very hierarchical''}, explained that leaders need to be involved in the process of bringing in new technology. They would have to understand its purpose and what benefits it can bring to their society: ``\textit{So the leadership would need to be able to be involved, evaluate and say `Okay'. As a community, this thing that is being used, uh, is it not something that is, uh, going to have a security implication to, to the village?}''

Other participants agreed with the involvement of the people in power and added that rural communities are open for technical improvements, but they want to make sure the technologies are introduced under their own terms, not in an imposed way:

``\textit{- you convince the top structures, then there's nothing that can stop you. In terms of accepting technology itself- the people are very mobile there.}''

Beyond appeasing authority, participants emphasized that this introductory process is necessary to ensure that drones and other new technologies are compatible with the cultural values of the community it is being introduced to: 

``\textit{It is their culture. Their perception. So you need to have a conversation with them and you also need to respect.}''

``\textit{You bring the technology ... not [as] a disruptive thing, so they're [compatible]  with the traditional hierarchy.}''

This understanding of emerging technologies as sociotechnical phenomenon that encompasses both people and technology was further stressed by one of the participants who stated the importance to clarify the role of new technologies in a historical as well as cultural context: 

\textit{``How does it relate to my history? Ancestral- you know?''}

Participants also stated the significance of history and tradition which cannot be forgotten in the design process. They acknowledged that different communities have particular taboos which need to be taken into consideration:

``\textit{You need to have a conversation with them and you also need to respect, um, certain areas that they might not want, um, the drone to have access to because they're still trying to protect, um, their culture and their heritage on certain things.}''

``\textit{So there would quite a lot of taboos that will be drawn. And then there would be areas where they'll be telling you, for example, this is an area where the-our ancestors are buried. And then you have your drone hovering there, then you are breaking taboos.}''

Finally, participants emphasized the importance of including the perspectives of people from rural communities throughout the entire design and deployment process. Advocating for a participatory approach, the participants described how stakeholders' opinions and suggestions should be listened to and understood. One participant described how he has been part of similar processes in the past: 

``\textit{I was part of a group going into the rural community to see how we can help to create a mini grid where there is no electricity. So we go to the community.}''

\subsubsection{Context-aware Drone Design}
A theme that emerged in the conversations connected the form and function of a drone closely to its context of use. The participants described how a drone's appearance can signify its role and functionality and should also be informed by the context in which it is used. Participants also mentioned differences between rural and urban communities as well as high diversities within different tribal areas: 

``\textit{The rural people may have a different design opinion...These are communities ... express certain things differently.}''

To illustrate, several participants described the significance of different colours in tribal communities: 

``\textit{So, for instance, in the design ... 
different colours may not be any important to the people who live in the city. But in the [rural] community, this is something that they care about.}''

Participants emphasized the importance of specific colours for different communities:

``\textit{You cannot wear red in our village. So red colour, you cannot even bring a red car, especially if it's in the rainy season. Then it's a taboo. Then, you know, you must take that into account.}''

``\textit{For instance, um, in Nigeria, some tribes, really, even in their cultural artifacts have a particular colour preference.}''

Participants linked these characteristics and the diversity of approaches towards them to a need for participatory approaches in designing and introducing new technologies in the rural communities: 

``\textit{I think that he needs to have the conversation with [the communities] and they'll be open to their ideas of also-- of what perceptions they have of different things. And then you incorporate that into the drone design. Because if you don't have the conversation, then you just create a drone with assumptions, then it's very different.}''

Beyond aesthetic qualities affecting form, participants described how functionality and perceived behavior should also take into account the context in which they are being developed: ``\textit{I wonder if the different context would require a different type of rule.}''

``\textit{If drones have emotions, it should depend on where they're used ... If they're being used on the homes they can have emotions ... in the areas let's say for security purposes in the context of agricultural and security in the funds they don't need to have emotions.}''

These comments show the high importance and relevance of the idea that the appearance -- look and feel -- of a drone can signify its role, and that an interpretation of the appearance can be culturally situated and depend on a community's particular customs or beliefs. 

\subsubsection{Data Ownership and Responsibility}
Another theme involved the ownership of data collected by drones and the responsibility that comes with it. Participants had different approaches towards this question. Several believed that ownership should be decided based on the power hierarchies already present in the communities. For example, one participant stated that in his community power structures were rigid, \textit{``especially in the rural areas because ... traditional leaders, from the chief, the village head, even at the family level, there are certain things you cannot do without involving the village head.''} Therefore, using drones and deciding who controls them and owns the data has to go through different levels of authority. Other participants described a more egalitarian approach that identified everyone as having some responsibility for the use of drones, \textit{``Everybody is responsible in the house, as long as it's the truth they should be able to have the same access.''}

One participant articulated a rationale for using an open access model to avoid misuse of the technology: ``\textit{I would take a more radical approach and say if you see it's a drone for community policing or community surveillance, then have it on open access for transparency's sake. A person with overall control, there's bound to be a misuse of whatever data is there. So you don't know, um, whether they gonna end up selling that data to, uh, the third party without your consent.}''

These different perspectives demonstrate a range of approaches towards questions of ownership and control when it comes to drones' data collection activities in different contexts.

\subsubsection{Human Surveillance Applications}
When discussing possible future applications, many expressed strong opinions about using drones for human surveillance. Participants had diverging approaches to this matter. Some stated that it would be desirable for drones to be able to detect peoples' presence and intentions, while others thought about this potential functionality as a transgression of their freedom. One participant proposed that the drone could use facial recognition to identify new people but refrain from deciding if they have good or bad intentions and postpone that judgment to a human actor. Such application could inform members of the community about newcomers: ``\textit{when the kids are playing, you can say we can identify new faces. It doesn't necessarily mean you can see new faces and [decide they are] a criminal. It could be a new child in the area, could be a new person, it could be whatever. So, like, if we detect them, the person who's responsible [can] just go and check.}''

Another participant was concerned about mistakes such technology could commit and misconceptions it could lead to: ``\textit{The moment you introduce ... being able to identify or assume intentions, there is a chance, I feel, that it might end up profiling a certain set of people or citizens who look a specific way, maybe based on my dressing.}''

These discussions expressed by the participants underline value tensions between security and privacy in this space. 

\subsubsection{Drone Applications}
Participants discussed possible future drone applications. Many of which incorporated geographical and environmental factors prevalent in sub-Saharan Africa. One participant described problems faced by rural communities during the rainy season, such as floods and the ease of getting lost. He suggested that drones could ``\textit{help people not to get lost in the field when going hunting. Especially if it's raining.}'' Further, he speculated that drones could be developed to ``\textit{protect yourself from lightning strikes.}''

Another participant mentioned the struggle of acquiring particular medications outside of urban areas. He described how drones could help deliver medicine from hospitals to smaller local clinics to \textit{``fast track ... send people ... different materials or medicine''} in the absence of good road infrastructures.

Finally, a participant explained that different drones should have various roles, and not all be responsible for the same tasks: ``\textit{You have a drone that is like a sheep dog that brings in ... the cows and sheep ... and then he secures them throughout the night; while the family drone would be entirely different, you know, helping the children to go to school.}''

Following this line of thought, another participant described how the appearance of a drone can signal its specific role, ``\textit{like wearing a uniform.}''

This notion of how the appearance of a drone corresponds to a specific role or application is corroborated in the findings of the design activity, presented in the following section.


\begin{figure}[t]
\includegraphics[width=8.5cm]{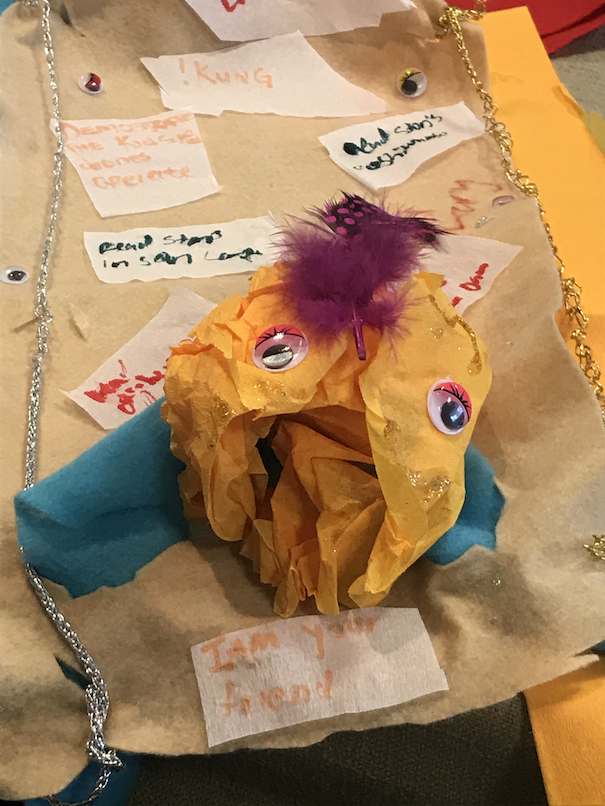}
\caption {Multi-lingual Library Drone which encourages people to read and can help them better pronounce and fix grammatical mistakes. This drone has many eyes to can see every corner of the library.}
\label{fig:Librarian}
\end{figure}

\subsection{Design Activity Findings}
Four drone designs were created for each application (station), with a total of twenty drones overall. The resulting designs were diverse in shape, form, and role given to the drone. We describe below the findings from the design activity.

\subsubsection{Applications and Use Cases}
For the same given application, the resulting drones presented diverse roles and use cases. For example, \textit{Agriculture} drones were designed to either: monitor aquatic life, watch over cattle and prevent lion attacks, detect plant diseases in farmed fields, or protect against theft. \textit{Security} drones were designed to monitor streets and detect criminals, record safety hazards, split fights between people, and protect cattle from wild animals. \textit{Learning} drones were designed as a school teacher, gym instructor, translator, and librarian (Figure \ref{fig:Librarian}). Finally, \textit{Entertainment} drones would take selfies, audio and video record events and extreme sports, or be used as party disco.

We found that the proposed use cases were both for outdoor (e.g., street, fields) and indoor (e.g., home, school) cases, for both city and rural environments, as well as for day and night usage. The chosen drone roles and tasks were based on participants' needs or the needs of the larger sub-Saharan population that they had previously encountered and/or identified. All drones were designed for positive activities with the intention to support and/or protect people, except the \textit{Entertainment} drones that had a more ``accessorized'' role. 
For example, one \textit{Learning} drone would encourage reading, another one would motivate students, and several drones would protect people and cattle against theft and animal attacks -- chasing lions --. The drones were primarily designed to support a community or a group of users ranging from children to elderly people, and from local people to clinics and police forces. None of the drones were designed for a single user, apart from one that was designed to \textit{``protect cattle without waking up the farmer.''} for which the farmer is potentially a single person. 

\begin{figure}[t]
\includegraphics[width=8.5cm]{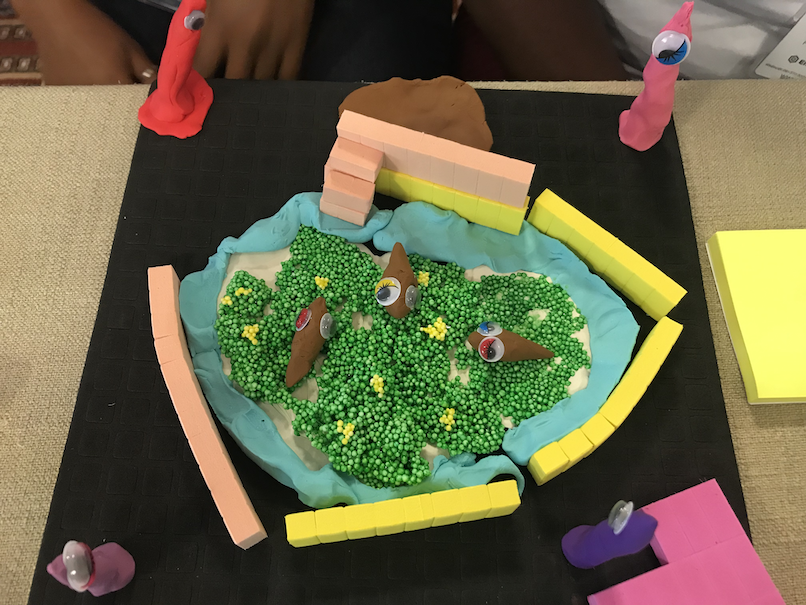}
\caption {This Drone Agricultural System patrols water resources. It has four surveillance drones which monitor the water coming from the dam, the surrounding areas, and the aquatic life inside the water.}
\label{fig:Dam}
\end{figure}

\subsubsection{Capabilities} \label{sec:autonomous}
The main capability that drones were given was the ability to take pictures and videos. As such, most drones (16) were designed with at least one camera. Interestingly, while about half of the drones (11) were given flying abilities, 9 were given ground features such as wheels, 5 had both flying and ground features, and one drone had the ability to both fly and swim.
Few drones were designed to move both on the ground and in the air. One main characteristic mentioned for the drones was the ability to understand native tribe and community languages in addition to English. 

The drones were primarily pictured as autonomous and able to make critical decisions, such as ``Miss Mouvel, a hospital assistant, responsible for communicating with patients who is able to prioritize patients by detecting the severity of pain.'' Few drones were designed as supporting tools, such as the \textit{Security} drone that can send information to the closest police station or the \textit{Healthcare} drone which delivers medication to the local clinic, allowing for anonymous healthcare. 

In some cases, participants were precise in the technical description, such as when discussing a drone equipped with GPS to not get lost, with a surveillance system, or mentioning ``its eyes are geared with computer vision to look around and make sense of the environment''. Two \textit{Healthcare} drones had drawers, including one where ``the medicine is locked in the special compartment in case someone would try to steal it''. One \textit{Security} drone could ``generate a fire flame and noise [...] to scare away predators'' and one \textit{Entertainment} drone could ``spin around like a disco ball and generate party smoke''.

\subsubsection{Interaction}
For most drones, participants mentioned some level of interaction between the drone and users. Most drones (12) were designed with apparent visual and/or audio feedback. For example, one drone used for medical transportation presented ``a complex cameras system and a display with a message to the recipient''. Only one drone (\textit{Entertainment}) proposed a specific interaction technique to take a selfie using a cuff that could be fitted on the user's wrist for manual control. The participants designed the cuff in addition to the drone to present their concept. Another (\textit{Entertainment}) drone was designed as an accessory ``flying selfie stick that can connect to the phone''. All other drones were independent. In most cases, communication and interaction was part of the drone's capabilities but was not explicitly formulated or designed, such as ``Miss Octopus'', a gym teacher drone, who ``motivates students with their swimming and running training'' or the \textit{Security} drone that can ``interact with citizens'' without further explanation. 

Across the designs, there was no mention of users' safety and no design choices related to safety mechanisms. It seemed obvious to participants that interaction was possible and safe, almost as a pre-requisite to the technology deployment. This is an interesting aspect since most commercial drones are not currently entirely safe for people around them. The only discussion around safety was about the drone's safety in the design of ``Miss Octopus'', who ``has tentacles which help her swim and ... protect her sensitive electronic parts.''

\begin{figure}[t]
\includegraphics[width=8.5cm]{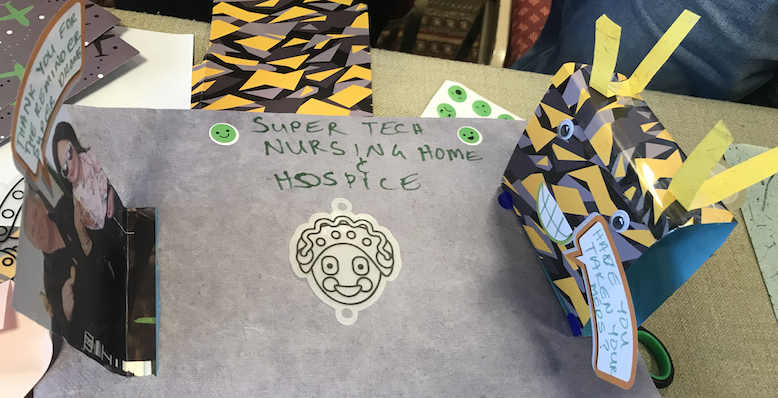}
\caption {``Super Tech Nurse Drone ... moves between the different rooms of the nursing home and reminds elderly people to take their medication. The speaker is hidden in the mouth and it has a camera in the eyes. It looks really friendly so that people will like it and trust it.''}
\label{fig:TechNurse}
\end{figure}

\subsubsection{Design}
Most drones (60\%) were created using several types of materials. 
Four were designed as part of a larger scene and not represented on their own, as to tell a story of how the drone would be used (Figures \ref{fig:Dam} \& \ref{fig:TechNurse}). One \textit{Agriculture} drone was designed as a swarm including several surveillance drones to monitor water coming from the dam (Figure \ref{fig:Dam}). 

When analyzing the designs, thirteen presented at least one element that can be recognized as a facial feature, all of them with eyes, and seven with a mouth. Eight drones had elements in their design that could be understood as limbs. As such, most drones presented either animal or human-like features. While one drone was designed to resemble an insect ``a bug that detects other bugs'' (Figure \ref{fig-teaser} Center), others were clearly anthropomorphized, not only designed with human-like features, but also described with human attributes. They were ``responsible'', with emotional characteristics, looking ``friendly'' and ``trustworthy''. They were even given names and assigned gender, such as ``Miss Mouvel'', ``she is a woman and she is friendly and caring.'' One \textit{Security} drone ``has a smiley face because it is proud of and happy with its job''.

In terms of shape, twelve drones had a body designed primarily in a square or cubic shape, with the remainder eight drones in a round or cylindrical shape. The \textit{Entertainment} drone ``Fleepy is small and has foldable wings, which makes it easy to put in the pocket and carry everywhere with you''. Colours played a role, such as the Librarian drone (Figure \ref{fig:Librarian}) which ``is really colourful to attract children.'' In several cases, different colours represented different aspects of the drone, such as blue representing water and red being connected to fire. In one case, the colours were chosen to ``resemble a police uniform so people will recognize it and trust it''.

\subsubsection{Best Drone Design Ratings}
All drones rated high on the -3 to 3 scale. The highest-rated design was the \textit{Agriculture} drone ``Jo-Ja'' (Figure \ref{fig-teaser} Center) with 2.62 (SD=0.65). The \textit{Learning} librarian drone (Figure \ref{fig:Librarian}) scored 2.46 (SD=0.66), the \textit{Healthcare} drone for medical deliveries rated 2.38 (SD=0.96), the \textit{Entertainment} party disco drone rated 2.23 (SD=1.09), and finally the \textit{Security} drone that monitors for safety hazards was rated 2.08 (SD=0.65).

\section{DISCUSSION}
Our results demonstrate the importance of working with community members to ensure that future drone designs address relevant real-world problems, benefits from the sociocultural perspectives and insights of diverse stakeholders, and is designed with human interactions in mind. In this section, we expand on and describe these outcomes in more detail.   


\subsection{Drones Solving Real-World Problems}
Across the workshop, drones were discussed and designed to specifically support needs and real problems that the participants or their communities face on a regular basis. The ideas generated would help alleviate some of the struggles, whether it is about providing multi-lingual services in a rich linguistic context\footnote{There are 27 individual living languages in Namibia. Of these, 22 are indigenous and 5 are non-indigenous.}, delivering medication to patients, chasing predators away from cattle, or preventing violence and theft. Many of the applications discussed during the workshop are specific to the local context and not previously presented in the scientific literature, highlighting the need to work with different regions of the world when designing new technologies.

In many cases, drones were designed and discussed in a supportive and protective role, acting as a kindergarten teacher, a nurse, a guide, a guard, or a police officer. While there was a tension between security and privacy during the focus group, the tension disappeared in the design activity. Indeed, when the drone was considered as its own autonomous being, as was the case for most designs (see \nameref{sec:autonomous}), the privacy concerns seemed to disappear, and we observed the drones being several times referred to as ``trustworthy''.    

One main topic addressed the contrast between rural and city living, which present different habits, infrastructures, and proximity to services. Participants discussed drone applications in both locations, yet, there was a strong emphasis on the support drones could provide rural communities. They highlighted problems, such as how roads might, in the wet season, be in such deteriorated condition that they cannot be used. Drone technology could already clearly positively impact these communities, ``leapfrogging infrastructure investment'' \cite{Sandvik2017}.

\subsection{Culturally-Situated Drone Design}
Much of the participants' input was centered around the importance of working directly with communities to inform the design of technologies intended for them. Participants described how different communities, even within the same country or region, hold values such as authority, privacy, and safety differently; and how designers need to incorporate, or at least be aware of these differences. These findings echo many of the principles underlying the co-design methodology that recognizes the importance of working with multiple stakeholders to ensure their perspectives are included in the design process. 

Most drone designs reflected the sociocultural characteristics of the participants' environments and cultural contexts. For example, participants discussed the lack of funds for people to travel to hospitals which are located in urban centers, and how this could be solved by having a drone delivering medication to clinics and patients. Furthermore, participants discussed how a drone could embed colours, behaviors, and features to represent values within a community. Another interesting finding was that the drones were discussed and designed to support a group of people and not a single user, while the large majority of commercial drones are designed for a single user. 

These findings are in par with earlier participatory design work in sub-Saharan Africa (e.g., \cite{Winschiers-Theophilus2010BeingApproach,Hamidi2018}). In particular, Winschiers-Theophilus et al. \cite{Winschiers-Theophilus2010BeingApproach} have discussed how community-based participatory approaches are rooted in traditional African cultures and can inform the design process. While we included some elements of storytelling and collaborative decision making in the current workshop, our approach was focused on gathering insights from diverse participants from different regions of Africa and facilitating co-design activities around a shared theme. In the future, research conducted with rural communities can benefit from using an approach similar to \cite{Winschiers-Theophilus2010BeingApproach} to further incorporate specific local and cultural elements. 

\subsection{Drones as Social Creatures}
The designed drones were predominantly anthropomorphized, presenting human features, characteristics, and emotional traits. They were given names and titles, some connected to the participants' culture, such as ``Teacha Lola'', and others associated to functionality, such as ``Master Translator''. The drones were described with specific behaviors, often protective and friendly, traits that were discussed in the focus group as community values that the drone could embed. One drone was, for example, explicitly given the ability to smile ``because it is proud of and happy with its job''.

In the focus group, the mention of a drone emotional response generated laughter. One participant then mentioned that emotions would make sense for drones used in homes but not for other applications such as security or agriculture. Yet, the designed drones were anthropomorphized to support acceptability and trust towards the technology, and described as intelligent and autonomous, capable of making complex and critical decisions, from hospital triage to taking care of children and elderly people. 
Most drones were given the ability to naturally and implicitly interact with people. This is in par with prior research on [ground] robotics which showed that people prefer human-like communication with social robots \cite{Dautenhahn2005WhatButler} and that anthropomorphism can be used to facilitate the acceptability of robots in natural human environments \cite{Zotowski2015:OppChall}. 

\section{Lessons Learned} 
Beyond the findings and discussion points presented above, we made several observations about the workshop organization and co-design activities that we present next. 

\subsection{Workshop Flexibility} 
We built a high-degree of flexibility into the workshop format. Participants would freely walk in and out of the room during the sessions, such as one participant who answered the phone during the focus group and left to take the call outside before joining back. Some participants chose to only take part in the focus group but not the design activity and vice versa. To cope with this, we ensured that, as people entered the room, they signed the consent form and filled in the surveys prior to joining the activity. They were then briefed on the current status of the workshop and tasks. In terms of time management, participants systematically asked for additional time to finish creating their design, so we added 5 min to the initial 15 min. Despite the workshop being a day long, all participants expressed wanting to complete all 5 stations. Because of time limitations, we moved on to the Idea Presentation stage after four design stations, as initially planned. We believe that having a fluid approach which consists in building in some degree of flexibility in workshops can make it easier for participants. This fits with Meyer's approach to scheduling in ``flexible time'' cultures \cite{meyer2014culture}. As such, participants can be included in the activities they would like to join - or can join - without feeling obligated to contribute to every stage of the process. 

\subsection{Moments of Celebration} Participants showed excitement and pride when presenting their designs. They took pictures and selfies with them, which they sent or posted on social media (e.g., Twitter). Some participants joined towards the end of the design activity after receiving a message from a friend or colleague who was taking part in the workshop. During the Idea Presentation stage, people who were not part of the workshop stepped in the room to see the final results. The atmosphere was loud, people were having fun, some were even cheering. Encouraging these dynamics might further motivate and encourage participants in similar workshops. 

\subsection{Physical Prototyping} In the design activity, even though pencils, pens, and crayons were provided in all stations, they were rarely used. Participants preferred to cut and fold 3D models instead of sketching. Even though all stations were well equipped with materials, participants would go to different stations to gather supplies, with a preference for eye stickers, glue, and scissors. In the fabric station, we wondered whether the participants would relate to the materials that were purchased from abroad, and which looked and felt differently from African textiles. Yet, this was not an issue and the drones were clearly designed for their needs regardless of the materials used.

\subsection{Owning Co-Design Activities} We found that participants appropriated the design activity to work on their favorite applications, partially regardless of the selected theme. For example, during the focus group, one of the applications discussed was \textit{wildlife monitoring}. However, at the application choosing stage, this application only received three votes and was not selected for the activity. Yet, two groups of participants used the \textit{Agriculture} station to create a drone that would support wildlife monitoring, highlighting their strong interest in this application.

\section{Future Work}
Our work touched on many topics from drone design to running co-design studies in sub-Saharan Africa. We address some of the interesting future work below. 

While prior work has shown that there is less focus on new hardware design for users in developing contexts \cite{Dell2016TheDevelopment}, our findings show the importance of including diverse stakeholder perspectives in hardware design to both ensure emerging technologies are sensitive to the needs of people in their context of deployment, and to enable contributions to the wider design and development communities. 

Our findings draw attention to the diversity of future applications and use cases, and to the general promise that drone technology could positively impact people in the region. Some uncovered use cases are not currently addressed in the literature, such as using drones to chase lions attacking cattle. We observed that drones were discussed as shared devices between groups of people, such as local communities, while drones are currently designed for a single person use. This finding has major implications on the interfaces to be developed for multi-user interaction strategy with drones but also in terms of ownership, such as when identifying who is responsible for the drone or for the data gathered by it. 

In the design activity, participants articulated the need for social features from the drones, and expressed instinctive trust in a potentially supportive and protective technology. We believe that different cultures may react differently to drone technology, bringing in different sets of expectations and concerns, and as such, future work is needed to fully appreciate cultural differences. We expect similar workshops will be run in other settings to further investigate different cultures and contexts.

\section{Conclusions}
We presented a co-design workshop for drone technology with participants from sub-Saharan Africa in Windhoek, Namibia. Participants described how the design of human-drone interaction should consider the social and cultural characteristics of its context of use and identified a series of applications reflecting the geographical and environmental factors of sub-Saharan regions. Participants also shared concerns about the privacy and regulation of drones, as well as the contrasting needs of future rural and urban users. This research addresses the need to include end-users' perspectives in the design of emerging technologies. In particular, it is the first work to include end-users in the co-design of drones, showing current missed opportunities for diversity and creativity that can arise from bringing in all stakeholders to the design table. Our findings validate the necessity of using participatory approaches for the design of drones in different contexts to ensure that they serve people's needs and remain contextually relevant. 

\section{Acknowledgments}
We would like to thank Matt Jones for his recommendations at the early stages of the project and the local team of AfriCHI 2018 in Windhoek, Namibia for enabling us to run this workshop and in particular to Anicia Peters for her hospitality during our stay. Our greatest thanks to the workshop participants for their openness and insights. 

\balance{}

\bibliographystyle{SIGCHI-Reference-Format}
\bibliography{biblio,magicLab_refs2}